\documentclass[aps,prb,floats,twocolumn,superscriptaddress]{revtex4}
\usepackage{amsfonts}
\usepackage{color}
\usepackage{enumerate}
\usepackage[utf8]{inputenc}
\usepackage{siunitx}
\usepackage{bm}
\usepackage{ dsfont }
\newcommand{\beq}{\begin{equation}}
\newcommand{\eeq}{\end{equation}}
\newcommand{\beqa}{\begin{eqnarray}}
\newcommand{\eeqa}{\end{eqnarray}}
\usepackage{graphicx,amssymb,amsmath,subfigure}
\newcommand{\ket}[1]{\left|#1\right>}

\begin{document}
\title{Magnetic field-Induced Nonlinear Optical Responses in Inversion Symmetric Dirac Semimetals}

\author{Alberto Cortijo}
\email{alberto.cortijo@csic.es}
\affiliation{Instituto de Ciencia de Materiales de Madrid,
CSIC, Cantoblanco; 28049 Madrid, Spain.}

\begin{abstract}
We show that, under the effect of an external magnetic field, a photogalvanic effect and the generation of second harmonic wave can be induced in inversion-symmetric and time reversal invariant Dirac semimetals and it is linear with the magnetic field. The mechanisms responsible of these non linear optical responses is the magnetochiral effect and the chiral magnetic effect.  What makes possible that these two effects give rise to the discussed non linear optical effects is the presence of band bending effects in the dispersion relation in real Dirac semimetals. Some observable consequences of this phenomenon are the appearance of a dc current on the surface of the system when it is irradiated with linearly polarized light or a rotation of the polarization plane of the reflected second harmonic wave.
\end{abstract}
\maketitle
\section{Introduction}
When a Weyl semimetal is placed under the effect of a magnetic field $\bm{B}$, the zeroth Landau level, characteristic of massless Dirac systems, splits up in two sectors of states with opposite chirality separated in momentum space. When an electric field $\bm{E}$ parallel to $\bm{B}$ is applied at the same time, this chirality valance breaks down, and the system becomes chiral. This is the origin of the chiral anomaly in Weyl semimetals\cite{NN83}.  It is now clear, as it can be seen employing semiclassical arguments for instance, that the chiral anomaly is a consequence of topologically non-trivial geometrical structures (the Berry curvature $\bm{\Omega}$ and the orbital magnetic moment $\bm{m}$\cite{XCN10}) appearing in Weyl and Dirac semimetals. The chiral anomaly is an observable phenomenon that leaves its imprint (in a direct and indirect way) in several transport and optical observables.

In the case of Dirac semimetals, like Cd$_3$As$_2$\cite{LJZ14,BGE14,NXS14,LGA15} or Na$_3$Bi\cite{LZZ14}, time reversal symmetry imposes that two sectors of states with opposite chirality lie around the same point in the Brillouin Zone, each sector belonging to a Karmers partner.  However, under the effect of electric and magnetic fields, the relative equilibrium between chiral states belonging to the same Kramers partner breaks down as in Weyl semimetals, and some consequences of the chiral anomaly that can be observed in Weyl semimetals appear in Dirac semimetals as well, like the $B^2$ dependent negative magnetoresistance\cite{NN83,SS13}. Other observables however, do not appear in Dirac semimetals. Inversion symmetry breaking\cite{HB12} can partially lift the constraint of having states of opposite chirality at the same point of the Brillouin Zone as it happens in TaAs\cite{WFF15,LXW15,YLS15} or TaP\cite{XWL16,ASW16}. Time reversal symmetry still forbids some manifestations of the chiral anomaly, like a quantum anomalous Hall current, but the breakdown of inversion symmetry now allows transport and/or optical effects related to the topological structures of the band structure that appear due to the lack of inversion symmetry\cite{MN16}. 
 
\begin{figure}
\includegraphics[scale=0.25]{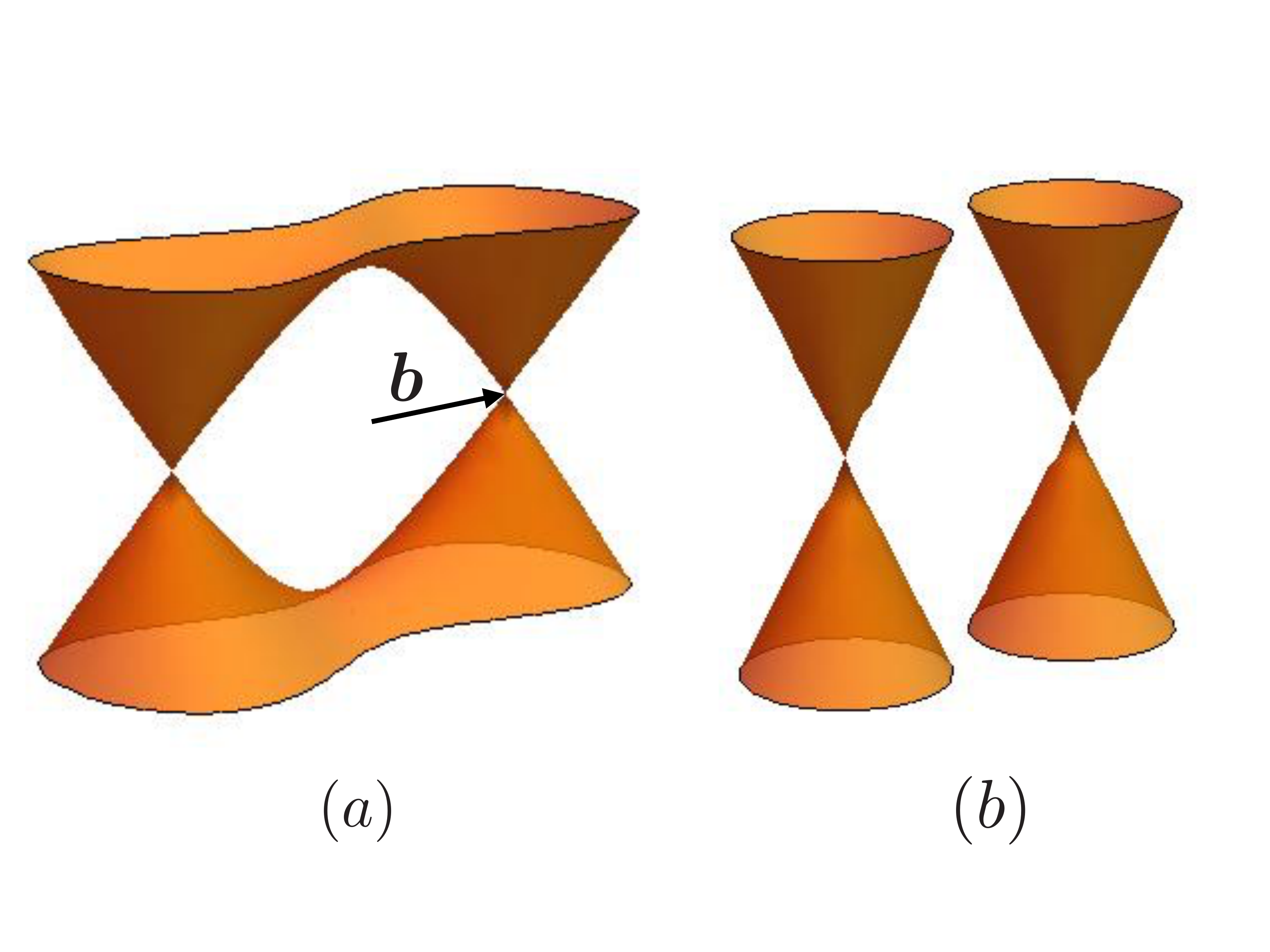}
\caption{(Color online) Bandstructure for Dirac semimetals. (a) represents the two-band model used in the text. The separation between nodes, the band bending, and the curvature in the dispersion relation is apparent. (b) represents the approximate linear dispersion relation around each Weyl node. The bands corresponding to the two Kramers partners are superimposed.}
\label{bands}
\end{figure}
From the perspective of Optics, from one side, it is known from long ago that chiral systems display (magnetic) optical activity, i.e., a difference in the response of the system to left- and right-polarized light proportional to the magnetic field. Also, the magnetic optical activity appears in non-chiral systems under the effect of an external magnetic field, appearing a similar difference between light polarizations. Then it is natural to expect a new response when a magnetic field is applied on a chiral system. This response, called magnetochiral effect, consists in a component of the dielectric tensor of the form $\varepsilon_{ij}\sim (\bm{B}\cdot\bm{k})\delta_{ij}$ and it was firstly discussed in the context of chiral molecular systems\cite{BBZ77,BZ79,RR97}.

From other side, natural optical activity, a phenomenon similar to the magnetic optical activity but with the momentum of the light playing the role of the magnetic field, ($\varepsilon_{ij}\sim \epsilon_{ijl}k_l$) has been predicted to occur in Weyl semimetals\cite{MP15,ZMS16} (but not in Dirac semimetals), appearing to be the same manifestation as the chiral magnetic effect\cite{FKW08}, when time dependent electromagnetic fields are considered. Since the natural optical activity is equivalent to the chiral magnetic effect in virtue of the Faraday's Law\cite{MP15}, it is hard to justify a term of the form $\bm{B}\cdot\bm{k}$ in the dielectric tensor, at least in the linear regime.
However, it has been recently proposed that a term similar to the magnetochiral effect should occur in Weyl semimetals in magneto-transport, where the separation between Weyl nodes $\bm{b}$ plays the role of the momentum vector\cite{A16}. From a naive perspective of the chiral anomaly in linearly dispersing Weyl systems, this term should not appear, since $\bm{b}$ acts as a chiral constant vector, and the chiral anomaly states that this vector can be almost gauged away with an appropriate chiral transformation, disappearing from the Weyl Hamiltonian but entering in the effective electromagnetic response of the system in the form of an axionic term\cite{W87,V03,B13}
\beq
\Delta\mathcal{S}=\frac{e^2}{8\pi^2}\int d^3\bm{x}dt (\bm{b}\cdot\bm{x})\bm{E}\cdot\bm{B},\label{axionlagrangean}
\eeq
that gives rise to the previously mentioned quantum anomalous Hall current $\bm{J}_{H}\sim\bm{b}\times\bm{E}$. The reason for this apparent discrepancy is that the model of fermions with (infinitely) linear dispersion relation is too restrictive, and band bending effects can enter in the observables associated to the topological structures. Also, the Onsager reciprocity relation allows for this effect since under time reversal symmetry, both $\bm{B}$ and $\bm{b}$ change their signs so the scalar product remains the same\cite{RF01}.

Contrary to the $B^2$ dependent negative magnetoresistance, that is quadratic in $\bm{\Omega}$, the (linear in $\bm{E}$ and $\bm{B}$) magnetochiral effect is linear in $\bm{\Omega}$ and $\bm{m}$, implying that it only appears in Weyl semimetals with broken time reversal symmetry.

We can then ask ourselves if the magnetochiral effect appears in non-linear optical responses in Dirac semimetals, since it is expected that part of these responses are quadratic in $\bm{\Omega}$ (and $\bm{m}$), implying that the contribution from Krarmers partners add up instead of cancelling. 
In the following sections we will use kinetic methods to compute the second order non-linear optical response (photogalvanic effect and second harmonic generation) of inversion-symmetric Dirac semimetals under the effect of a magnetic field, showing that these responses are is indeed a non zero and are proportional to the product $\bm{B}\cdot\bm{b}$.

Besides, obtaining a non vanishing second non-linear optical response in inversion and time reversal symmetric systems in bulk is valuable by its own right, even if these responses depend on the application of an external magnetic field, because the magnetic field is even under inversion, and it is not enough to trigger such responses. 

The paper is organized as follows: In section \ref{sec:model} we highlight the important features of the model used an how it is related to the well known Weyl Hamiltonian. In section \ref{sec:kinetics} the derivation of the kinetic equations for the non linear terms and the current is presented. In section \ref{sec:allPGE} we describe the photogalvanic effect depending on the type of irradiating light while in section \ref{sec:allSHG} we do the same for the second harmonic generation. We finally discuss these results in section \ref{sec:discussion}. 
\section{The model}
\label{sec:model}
In the case of dealing with Dirac semimetals, the two Kramers partners must be taken into account. Also, if inversion symmetry is required, the two Kramers partners are superimposed to each other. We thus consider the following two-band Hamiltonian in the continuum\cite{LZS15,CY15b,SHS16,A16}:
\beq
H(\bm{k})=s v\bm{\sigma}_{\perp}\cdot\bm{k}_{\perp}+s\sigma_3(m-\beta k^2_3),\label{Ham0}
\eeq
that generalizes the Hamiltonian for linearly dispersing states of chirality $\chi=\pm1$: $H=\sum_s s\chi v\bm{\sigma}\cdot(\bm{k}-\chi \bm{b})$. The index $s=\pm 1$ stands for the two Kramers partners in $\mathcal{T}$ symmetric Dirac semimetals. Throughout the paper we will set $\hbar=c=1$ for simplicity. It is not hard to recover them in the final expressions.

For each value of $s$, the spectrum described by (\ref{Ham0}) when $m\cdot\beta>0$ consists in two bands touching at two Weyl points placed at $\bm{b}_{\chi}=(\bm{0}_{\perp},\chi b_3)$, with $b_3=\sqrt{m/\beta}$.

Expanding around these two Weyl nodes we have ($v_3=2\sqrt{m\beta}$):
\beq
H(\bm{b}_{\chi}+\delta\bm{k})=s v\bm{\sigma}_{\perp}\cdot\delta\bm{k}_{\perp}-s\sigma_3\left(\chi v_3\delta k_3+\frac{v_3}{2b_3} \delta k^2_3\right).\label{Ham01}
\eeq
The bandstructure around these two Weyl points describes two massless linearly dispersing fermions with opposite chirality (isotropic when $v_3= v$), when we can drop the quadratic term in $\delta k_3$ in (\ref{Ham01}). The difference between the spectrum obtained by the model (\ref{Ham0}) and the purely linear spectrum can be seen in Fig. \ref{bands}. If we want to recover a linear spectrum with a finite $v_3$, in practice we have to consider that $b_3$ should be much larger than any other energy/momentum scale, like the Fermi level $\mu$. This implies that we can recover the results that one would obtain with the linear model in the results obtained here by taking the limit $b_3\rightarrow \infty$. The advantage of the model (\ref{Ham0}) over the linear model is that we are able to explore situations where $b_3$ is not large, or the Fermi level is above (or below) the Van Hove point at $|m|$ ($-|m|$). In real Dirac semimetals like alkali-doped Cd$_3$As$_2$, the Fermi wavenumber $k_F$ and the momentum $b_3$ might be of the same order (or even larger) depending on the induced doping\cite{LJZ14}.

\section{Non linear magneto-optics in kinetic theory}
\label{sec:kinetics}
In this section we will adapt the semiclassical theory of non-linear magnetooptics to the model at hands\cite{DGI09,MZO16,EFO16,TIS16,MN16}. Without loss of generality, we will assume that the Fermi level $\mu$ crosses the conduction band. We will consider the photogalvanic effect (PGE) and second harmonic
generation (SHG) in a Dirac semimetal for homogeneous and static magnetic fields $\bm{B}$, and homogeneous and time dependent electric fields:
\beq
\bm{E}(t)=\bm{E}_1\cos(\omega t)+\bm{E}_2\sin(\omega t),\label{Efield}
\eeq
or, in complex notation
\beq
\bm{E}(t)=\bm{\mathcal{E}} e^{-i \omega t}+\bm{\mathcal{E}}^{*}e^{i\omega t},\label{Efieldcomplex}
\eeq
with $\bm{\mathcal{E}}=(\bm{E}_1+i\bm{E}_2)/2$, and $\bm{\mathcal{E}}^{*}$ the complex conjugate
of $\bm{\mathcal{E}}$.

The equations of motion for $\dot{\bm{k}}$ and $\dot{\bm{x}}$ are\cite{XCN10}:
\begin{subequations}
\beq
D \dot{\bm{k}}=e\bm{E}+e \bm{v}\times \bm{B}+e^2(\bm{E}\cdot \bm{B})\bm{\Omega},\label{forceeq}
\eeq
\beq
D\dot{\bm{x}}=\bm{v}-e\bm{\Omega}\times\bm{E}+e(\bm{v}\cdot\bm{\Omega})\bm{B},\label{velocityeq}
\eeq
\end{subequations} 
where $\bm{v}=\frac{\partial\epsilon(\bm{k})}{\partial \bm{k}}$ is the group velocity with the dispersion relation modified by the orbital magnetic moment $\bm{m}(\bm{k})$: $\epsilon(\bm{k})=\epsilon_0(\bm{k})-e\bm{m}(\bm{k})\cdot\bm{B}$. The coefficient $D=1+e\bm{\Omega}\cdot \bm{B}$ is the volume element of the phase space. The vector $\bm{\Omega}(\bm{k})$ is the Berry curvature of the considered band.

We have to solve the time dependent Boltzmann equation for the non-equilibrium distribution function $f(t,\bm{k})$ adopting the relaxation time approximation:
\beq
\frac{\partial f(t,\bm{k})}{\partial t}+\dot{\bm{k}}\cdot\frac{\partial f(t,\bm{k})}{\partial \bm{k}}=-\frac{1}{\tau}\left(f(t,\bm{k})-f_{0}(\bm{k})\right).\label{boltzmanntot}
\eeq
The parameter $\tau$ is the transport time, and $f_0(\bm{k})=f_0(\epsilon_0(\bm{k}))$ is the equilibrium Fermi-Dirac distribution in \emph{absence} of any external field.
Since we are interested in non linear terms, quadratic in the electric field $\bm{E}$, we will recursively solve the equation (\ref{boltzmanntot}) assuming that $f(t,\bm{k})\simeq f_0+f_1+f_2$, with
$f_1\sim O(E)$ and $f_2 \sim O(E^2)$\cite{SF15}. We have to substitute this ansatz in (\ref{boltzmanntot}) and obtain the kinetic equations for $f_{1}$ and $f_{2}$ retaining the corresponding powers of the electric field.

Finally, it is important to keep in mind that we are interested in the effect of the magnetic field $\bm{B}$ on the quasiparticle current to linear order, so we will solve the resulting kinetic equations up to first order in $\bm{B}$ by using the Jones-Zener method\cite{JZ34}. Higher orders in the magnetic fields can be systematically computed using this method. 
\subsection{Kinetic equation for $f_1$}
If we multiply the equation (\ref{boltzmanntot}) by $D$ and keep linear orders in the electric field, we have

\beqa
\tau D\frac{\partial f_1}{\partial t}+Df_1+\tau D\dot{\bm{k}}\cdot\frac{\partial f_1}{\partial \bm{k}}=-\tau D \dot{\bm{k}}\cdot\bm{v}_0\frac{\partial f_0}{\partial \epsilon_{\bm{k}}},\label{boltzmann1}
\eeqa
after using that $\frac{\partial f_0}{\partial \bm{k}}=\bm{v}_0\frac{\partial f_0}{\partial \epsilon_{\bm{k}}}$.

To linear order in $\bm{B}$, the product $\tau D\dot{\bm{k}}\cdot\partial/\partial \bm{k}$ reads
\beqa
\tau D\dot{\bm{k}}\cdot\frac{\partial}{\partial\bm{k}}&=&\tau e (\bm{v}_0\times\bm{B})\cdot\frac{\partial}{\partial\bm{k}}+\nonumber\\
&+&\tau e\bm{E}\cdot\frac{\partial}{\partial\bm{k}}+\tau e^2(\bm{E}\cdot\bm{B})\bm{\Omega}\cdot\frac{\partial}{\partial\bm{k}}.\label{Boperator}
\eeqa
the first term in (\ref{Boperator}) is order $O(E^0)$ while the two last terms are $O(E)$. It means that, to order $O(E)$, only the first term of the right hand side has to be kept in eq.(\ref{boltzmann1}).

In the case of the term $\tau D \dot{\bm{k}}\cdot\bm{v}_0$, we have
\beqa
\tau D \dot{\bm{k}}\cdot\bm{v}_0=\tau e \bm{E}\cdot\bm{v}_0+\tau e^2(\bm{E}\cdot\bm{B})\bm{\Omega}\cdot\bm{v}_0.\label{independentterm}
\eeqa
Here it is important to note that both terms are order $O(E)$.

Since $f_1$ is $O(E)$, the form of the electric field (\ref{Efield}) suggests to write $f_1$ as $f_1(t,\bm{k})=f_1(\bm{k})e^{-i\omega t}+f^{*}_1(\bm{k})e^{i\omega t}$ ($f^{*}_1$ being the complex conjugate of $f_1$). Substituting the expressions (\ref{Boperator}) and (\ref{independentterm}) in (\ref{boltzmann1}), and expanding $1/D$ to linear order in $\bm{B}$, we get

\beqa
&&\alpha_1 f_1+\tau e (\bm{v}\times \bm{B})\cdot\frac{\partial f_1}{\partial \bm{k}}=-\tau e\frac{\partial f_0}{\partial \epsilon_{\bm{k}}}\bm{\mathcal{E}}\cdot \bm{v}_0+\nonumber\\
&+&\tau e^2\frac{\partial f_0}{\partial \epsilon_{\bm{k}}}(\bm{B}\cdot\bm{\Omega})\bm{\mathcal{E}}\cdot\bm{v}_0-\tau e^2\frac{\partial f_0}{\partial \epsilon_{\bm{k}}}(\bm{\bm{\mathcal{E}}}\cdot\bm{B})
\bm{\Omega}\cdot\bm{v}_0,\label{boltzmann1F}
\eeqa
with the phase factor $\alpha_1=1-i\omega\tau$.

Now we take advantage of the choice $\bm{B}||\bm{b}$, and that the model is axisymmetric along the direction $\bm{b}$:
\beq
\tau e(\bm{v}_0\times \bm{B})\cdot\frac{\partial}{\partial \bm{k}}=-\tau e B\frac{v^2}{\epsilon_{\bm{k}}}\frac{\partial}{\partial \theta},
\eeq
where $\theta$ is the polar angle defined in the plane perpendicular to $\bm{B}$.

Dividing the equation (\ref{boltzmann1F}) by the phase factor $\alpha_1$, defining the operator $\hat{\Theta}_1=-\frac{\tau}{\alpha_1}eB\frac{v^2}{\epsilon_{\bm{k}}}\partial_{\theta}$, and noticing that the operator $\hat{\Theta}_1$ is linear in the magnetic field, we can formally write $f_1(\bm{k})$ as the result of applying the integral operator $(1+\hat{\Theta}_1)^{-1}$ to the right hand side of (\ref{boltzmann1F}). Writing this integral operator  as an infinite series of $\hat{\Theta}_1$ and to linear order in the magnetic field we get:
\beqa
f_1&=&-\frac{\tau}{\alpha_1} e\frac{\partial f_0}{\partial \epsilon_{\bm{k}}}\bm{\mathcal{E}}\cdot \bm{v}_0+\frac{\tau^2}{\alpha^2_1}e^2(\bm{v}_0\times \bm{B})\cdot\frac{\partial (\bm{v}_0\cdot\bm{\mathcal{E}})}{\partial \bm{k}}-\nonumber\\
&-&\frac{\tau}{\alpha_1} e^2\frac{\partial f_0}{\partial \epsilon_{\bm{k}}}(\bm{\mathcal{E}}\cdot\bm{B})\bm{\Omega}\cdot\bm{v}_0+\frac{\tau}{\alpha_1} e^2\frac{\partial f_0}{\partial \epsilon_{\bm{k}}}(\bm{B}\cdot\bm{\Omega})\bm{\mathcal{E}}\cdot\bm{v}_0\equiv\nonumber\\
&\equiv&\bm{\Delta}^{(0,0)}\cdot\bm{\mathcal{E}}+\bm{\Delta}^{(1,0)}\cdot\bm{\mathcal{E}}+\bm{\Delta}^{(1,1)}\cdot\bm{\mathcal{E}}.\label{solboltzmann1}
\eeqa
In equation (\ref{solboltzmann1}), the vectors $\bm{\Delta}^{(m,n)}$ represent the parts of the distribution $f_1$ that are powers of $\bm{B}^m$ and $\bm{\Omega}^n$. The part corresponding to $f^{*}_1$ is obtained by computing the complex conjugate of $f_1$.
The same expression has been recently obtained in the static limit ($\omega\rightarrow 0$) in the context of the imprint of the magnetochiral effect in the magnetotrasport in WSM\cite{A16}.
\subsection{Kinetic equation for $f_{20}$}

As we have mentioned before, the distribution function $f_2$ depends quadratically of the electric field. Since we are using complex notation, by using (\ref{Efieldcomplex}), one gets two contributions to $f_2$: a term $f_{20}$ coming from the product of $\bm{\mathcal{E}}$ and its complex conjugate $\bm{\mathcal{E}}^*$, that does not depend on time, and a component $f_2$ coming from the product $\bm{\mathcal{E}}\cdot\bm{\mathcal{E}}$ (and its complex conjugate $\bm{\mathcal{E}}^*\cdot\bm{\mathcal{E}}^*$) that depends on time as $e^{2i\omega t}$. Both terms have different physical consequences, as it is well known.

To linear order in $\bm{B}$, the kinetic equation for $f_{20}$ reads
\beqa
&&f_{20}+\tau e(\bm{v}_0\times\bm{B})\cdot\frac{\partial f_{20}}{\partial \bm{k}}=-\tau e\bm{\mathcal{E}}^*\cdot\frac{\partial f_1}{\partial\bm{k}}+\nonumber\\
&+&\tau e^2(\bm{B}\cdot\bm{\Omega}) \bm{\mathcal{E}}^*\cdot\frac{\partial f_1}{\partial\bm{k}}-\tau e^2(\bm{\mathcal{E}}^*\cdot\bm{B})\bm{\Omega}\cdot\frac{\partial f_1}{\partial\bm{k}}+\nonumber\\
&+&c.c.\label{boltzmann2}
\eeqa

As we did for $f_1$, we can define the operator $\hat{\Theta}_0=-\tau eB\frac{v^2}{\epsilon_{\bm{k}}}\partial_{\theta}$ and write $f_{20}$ as powers series of the operator $(1+\hat{\Theta}_0)^{-1}$ applied to the right hand side of (\ref{boltzmann2}). To linear order in $\bm{B}$, we have, in terms of $f_1$:
\beqa
f_{20}&=&-\tau e\bm{\mathcal{E}}^*\cdot\frac{\partial f_1}{\partial\bm{k}}+\tau e^2(\bm{B}\cdot\bm{\Omega}) \bm{\mathcal{E}}^*\cdot\frac{\partial f_1}{\partial\bm{k}}-\nonumber\\
&-&\tau e^2(\bm{\mathcal{E}}^*\cdot\bm{B})\bm{\Omega}\cdot\frac{\partial f_1}{\partial\bm{k}}+\nonumber\\
&+&\tau^2 e^2(\bm{v}_0\times\bm{B})\cdot\frac{\partial }{\partial \bm{k}}\left(\bm{\mathcal{E}}^*\cdot\frac{\partial f_1}{\partial\bm{k}}\right)+c.c.\label{solboltzmann2}
\eeqa
Now, we have to insert in (\ref{solboltzmann2}) the expression (\ref{solboltzmann1}) for $f_1$, keeping the linear terms in $\bm{B}$.
\subsection{Kinetic equation for $f_{2}$}
To linear order in $\bm{B}$, the kinetic equation for $f_{2}$ reads
\beqa
&&\alpha_2 f_{2}+\tau(\bm{v}_0\times\bm{B})\cdot\frac{\partial f_{2}}{\partial \bm{k}}=-\tau e\bm{\mathcal{E}}\cdot\frac{\partial f_1}{\partial\bm{k}}+\nonumber\\
&+&\tau e^2(\bm{B}\cdot\bm{\Omega}) \bm{\mathcal{E}}\cdot\frac{\partial f_1}{\partial\bm{k}}-\tau e^2(\bm{\mathcal{E}}\cdot\bm{B})\bm{\Omega}\cdot\frac{\partial f_1}{\partial\bm{k}}+\nonumber\\
&+&c.c.\label{boltzmann22}
\eeqa
in this case, we have defined the phase factor $\alpha_2=1-2i\omega\tau$. The solution is similar to (\ref{solboltzmann2}) with the appropriate changes:
\beqa
f_{2}&=&-\frac{\tau}{\alpha_2} e\bm{\mathcal{E}}\cdot\frac{\partial f_1}{\partial\bm{k}}+\frac{\tau}{\alpha_2} e^2(\bm{B}\cdot\bm{\Omega}) \bm{\mathcal{E}}\cdot\frac{\partial f_1}{\partial\bm{k}}-\nonumber\\
&-&\frac{\tau}{\alpha_2} e^2(\bm{\mathcal{E}}\cdot\bm{B})\bm{\Omega}\cdot\frac{\partial f_1}{\partial\bm{k}}+\nonumber\\
&+&\frac{\tau^2}{\alpha^2_2} e^2(\bm{v}_0\times\bm{B})\cdot\frac{\partial }{\partial \bm{k}}\left(\bm{\mathcal{E}}\cdot\frac{\partial f_1}{\partial\bm{k}}\right)+c.c.\label{solboltzmann22}
\eeqa
\subsection{Quasiparticle current in the local limit}
After computing all the relevant components of the non-equilibrium distribution function $f$, one then plugs it and velocity $\dot{\bm{x}}$ defined in eq.(\ref{velocityeq}) into the quasiparticle current $\bm{J}=e\int (d\bm{k})f(t,\bm{k})D\dot{\bm{x}}$ to get ($(d\bm{k})\equiv\frac{d^3\bm{k}}{8\pi^3}$):
\beqa
\bm{J}&=&e\int (d\bm{k})(f_1+f_2)(\bm{v}_0-e\bm{\Omega}\times\bm{E}-\nonumber\\
&-&e\frac{(\partial \bm{m}\cdot\bm{B})}{\partial \bm{k}}+e(\bm{v}_{0}\cdot\bm{\Omega})\bm{B}).\label{current}
\eeqa
For the part of the current linear in the electric field, we have
\beqa
\bm{J}^{(1)}&=&e\int (d\bm{k})f_1(\bm{v}_0-e\frac{(\partial \bm{m}\cdot\bm{B})}{\partial \bm{k}}+\nonumber\\
&+&e(\bm{v}_{0}\cdot\bm{\Omega})\bm{B}),
\eeqa
while, by simple inspection, we see that the second order non linear current $\bm{J}^{(2)}$ will receive two contributions. The first comes from the the combination of $f_1$ and the anomalous velocity term:
\beq
\bm{J}^{(2,1)}=-e^2\int (d\bm{k}) f_1\bm{\Omega}\times\bm{E},\label{current21}
\eeq
while the second contribution comes from $f_2$ and the remaining terms of $\dot{\bm{x}}$,
\beq
\bm{J}^{(2,2)}=e\int (d\bm{k}) f_2(\bm{v}_0-e\frac{\partial (\bm{m}\cdot\bm{B})}{\partial \bm{k}}+e(\bm{v}_{0}\cdot\bm{\Omega})\bm{B}).\label{current22}
\eeq
Since $\bm{\Omega}^s=s\bm{\Omega}$, and $\bm{m}^s=s\bm{m}$ with $s=\pm1$ labeling the two Kramers partners, for time reversal invariant Dirac semimetals, only the terms of $\bm{J}$ that do not depend on $\bm{\Omega}$ and $\bm{m}$, or are quadratic in $\bm{\Omega}$ and $\bm{m}$ or proportional to the product $\bm{\Omega}\cdot\bm{m}$ will survive when summing the currents for both partners: $\bm{J}=\sum_{s}\bm{J}_{s}$.

\section{Photogalvanic Effect in Dirac semimetals}
\label{sec:allPGE}
 To linear order in the magnetic field, it is apparent by inspection that the only terms that do not depend on $\bm{\Omega}$ or $\bm{m}$ or are quadratic in
 them are
\beq
\bm{J}^{(2,1)}_1=-2e^2\int (d\bm{k})(\bm{\Delta}^{(1,1)}\cdot\bm{\mathcal{E}})\bm{\Omega}\times\bm{\mathcal{E}}^{*}+c.c,\label{1contribution0}
\eeq
which is quadratic in powers of the Berry curvature $\bm{\Omega}$ and
\beqa
&&\bm{J}^{(2,2)}_2=-\tau e^2\int (d\bm{k})\bm{v}_0\bm{\mathcal{E}}^*\cdot\frac{\partial}{\partial \bm{k}}(\bm{\Delta}^{(0,0)}\cdot\bm{\mathcal{E}})-\nonumber\\
&-&\tau e^2\int (d\bm{k})\bm{v}_0\bm{\mathcal{E}}^*\cdot\frac{\partial}{\partial \bm{k}}(\bm{\Delta}^{(1,0)}\cdot\bm{\mathcal{E}})+\nonumber\\
&+&\tau^2 e^3\int (d\bm{k})\bm{v}_0(\bm{v}_0\times\bm{B})\cdot\frac{\partial}{\partial \bm{k}}(\bm{\mathcal{E}}^*\cdot\frac{\partial}{\partial \bm{k}}(\bm{\Delta}^{(0,0)}\cdot\bm{\mathcal{E}}))+\nonumber\\
&+&c.c.\label{2contribution}
\eeqa

which is the part of the current that does not depend on $\bm{\Omega}$ or $\bm{m}$. Remarkably, despite the complexity of the expression (\ref{2contribution}), for inversion symmetric systems like the ones considered here, this contribution vanishes\cite{BS80}. It can be seen in two steps: first we can integrate eq.(\ref{2contribution}) by parts and neglect surface terms. Then, it is enough to notice that, for inversion-symmetric systems, the velocity $\bm{v}^0$ and the derivative $\partial_{\bm{k}}$ change sign under inversion. 

It means that the only photogalvanic current comes from the contribution (after using the definition of $\bm{\Delta}^{(1,1)}$ in (\ref{solboltzmann1}))
\beqa
\bm{J}^{PGE}_1&=&-2e^4\frac{\tau}{\alpha_1}\int (d\bm{k})\frac{\partial f_0}{\partial \epsilon_{\bm{k}}}
[(\bm{\Omega\cdot\bm{B}})(\bm{\mathcal{E}}\cdot\bm{v}_0)-\nonumber\\
&-&(\bm{B}\cdot\bm{\mathcal{E}})(\bm{\Omega}\cdot\bm{v}_0)]\bm{\Omega}\times\bm{\mathcal{E}}^*+c.c.,\label{PGEcurrent}
\eeqa 
or, in components,
\beqa
J^{PGE}_a&=&-2e^4\frac{\tau}{\alpha_1}\epsilon_{abc}\int (d\bm{k})\frac{\partial f_0}{\partial \epsilon_{\bm{k}}}\Omega_b\Omega_r\cdot\nonumber\\
&&\cdot[B_r v^{0}_d-B_dv^{0}_r]\mathcal{E}^*_c\mathcal{E}_d+c.c.
\eeqa
If we assume zero temperature and choose the magnetic field to be parallel to $\bm{b}$, the integral in $\bm{k}$ can be performed obtaining, in the regime $\mu\ll v b_3$,
\beq
J^{PGE}_a=\left(\frac{e^4B_3b_3\tau v}{30\pi^2 \alpha_1\mu b^2_3}\right)\epsilon_{abc}\Lambda_{bd}\mathcal{E}^*_c\mathcal{E}_d+c.c.,\label{PGEcurrentcomponents}
\eeq
where we have defined the diagonal matrix $\Lambda_{bd}=diag(-1,-1,2)$. One sees that the effect is proportional to $B_3b_3$ (for the electromagnetic configuration chosen), implying that it is the magnetochiral effect what is behind of the appearance of a magneto-photogalvanic current in inversion symmetric Dirac semimetals. It might seem not apparent that the dependence $B_3b_3$ is equivalent to write the scalar product $\bm{B}\cdot\bm{b}$. It was shown in ref.\cite{A16} that this is indeed the case.

Alternatively, we can choose the magnetic field to be perpendicular to $\bm{b}$: $\bm{B}\perp\bm{b}$. In this case, the magnetochiral effect is zero, but there is still a contribution to the current (\ref{PGEcurrent}):
\beq
\bm{J}^{PGE}=\left(\frac{11e^4\tau v}{30\pi^2 \alpha_1\mu b^2_3}\right)(\bm{B}_{\perp}\cdot\bm{\mathcal{E}})\bm{b}\times\bm{\mathcal{E}}^{*}+c.c.\label{PGEAxialanomaly}
\eeq
This part of the current is now proportional to the product $\bm{B}\cdot\bm{\mathcal{E}}$ coming from the chiral magnetic effect term in the equation of motion (\ref{velocityeq}). It is interesting to note that the product $\bm{b}\times\bm{\mathcal{E}}^{*}$ is similar in form to the natural optical activity in dielectrics: $\varepsilon_{ij}\sim\bm{k}\times\bm{\mathcal{E}}^*$ that has been discussed in time reversal breaking Weyl semimetals\cite{MP15,ZMS16}.

As we discussed in the previous sections, we can obtain from (\ref{PGEcurrentcomponents}) and (\ref{PGEAxialanomaly}) the result that one would get if the model of linearly dispersing Weyl fermions were used. Taking the limit of $b_3$ very large, one obtains that $\bm{J}^{PGE}$ is zero in this limit, concluding that the photoinduced ac currents (\ref{PGEcurrentcomponents}) and (\ref{PGEAxialanomaly}) is a consequence of the presence of curvature effects in the bandstructure.
\begin{figure*}
\begin{minipage}{.47\linewidth}
(a)
\includegraphics[scale=0.22]{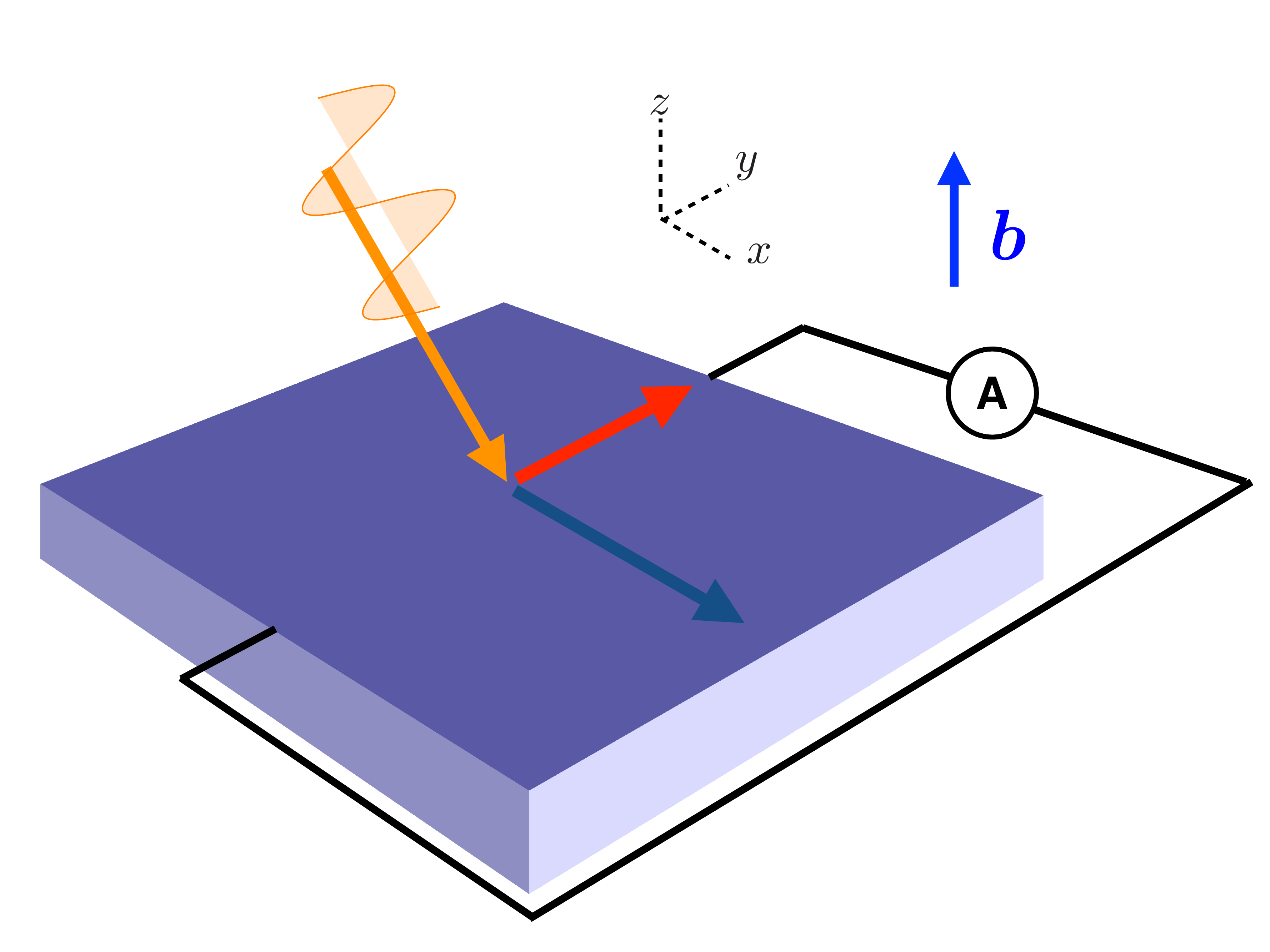}
\end{minipage}
\begin{minipage}{.48\linewidth}
(b)
\includegraphics[scale=0.4]{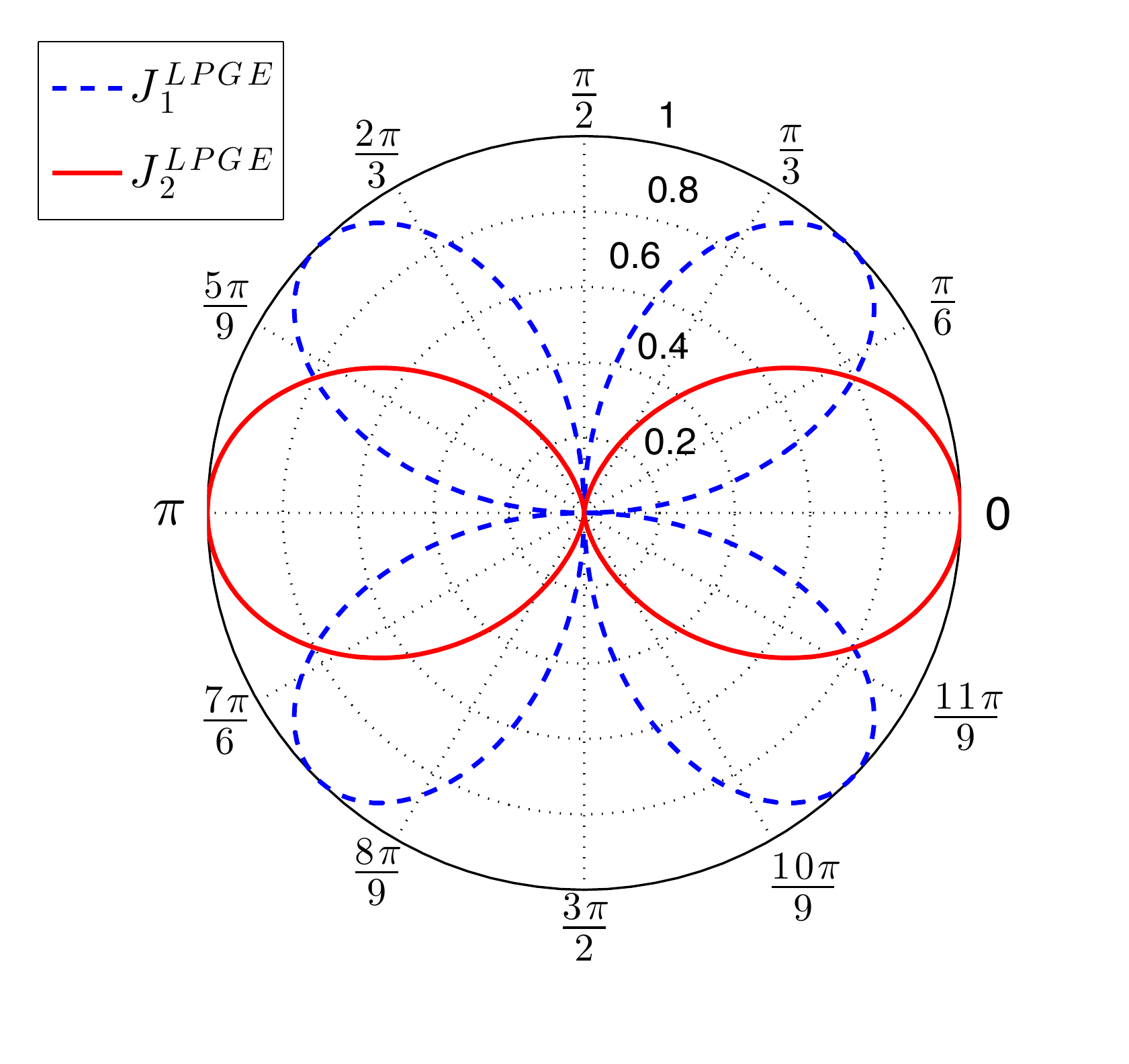}
\end{minipage}
\caption{(Color online)(a) Prototypical experimental setup employed to measure the surface current generated by linearly polarized light. A second amperemeter can be used to simultaneously measure the other component of the current. (b) Angular dependence of the components of the induced current $\bm{J}^{LPGE}$ as a function of the polarization angle $\alpha$ of the incident light.}
\label{Figs2}
\end{figure*}
\subsection{Linear PGE at wave incidence $\theta_i$ and polarization angle $\alpha$}
Here we will consider the PGE under the influence of a plane wave linearly polarized of polarization $\alpha$ at incidence angle $\theta_i$, and the magnetic field $\bm{B}$ parallel to $\bm{b}$. As we have discussed previously, in this situation the PGE current comes from the magnetochiral effect. The plane of incidence is chosen to be $y=0$. Linearly polarized waves are real so $\bm{\mathcal{E}}=\bm{\mathcal{E}}^*$. We will consider the following form for the electric field:
\beqa
\bm{\mathcal{E}}=\frac{E_0}{2}\left(\begin{array}{c}
\cos\theta_i \cos\alpha\\
\sin\alpha\\
\sin\theta_i \cos\alpha
\end{array}\right).
\eeqa
$E_0$ is the intensity of the electric field. Inserting this expression in (\ref{PGEcurrentcomponents}), we obtain
\beqa
\bm{J}^{LPGE}=\frac{\tau e^4B_3 vE^2_0}{40\pi^2\mu b_3(1+\omega^2\tau^2)}\left(\begin{array}{c}
-\sin\theta_i \sin2\alpha\\
\sin2\theta_i\cos^2\alpha\\
0
\end{array}\right).\label{LPGC}
\eeqa
Under the effect of a linearly polarized electromagnetic wave, an in-plane dc current (\ref{LPGC}) is generated for these samples where the vector $\bm{b}$ is normal to the surface. The direction of the current can be controlled by the polarization of the incident wave. The angular dependence of the components of $\bm{J}^{LPGE}$ with the polarization angle are plotted in Fig.\ref{Figs2}b. A typical experimental setup employed\cite{OYY16} to measure such photocurrents is plotted in Fig.\ref{Figs2}a. 

When the magnetic field is perpendicular to $\bm{b}$, the current (\ref{PGEAxialanomaly}) gives (we have chosen $\bm{B}=B_1\hat{\bm{x}}$ for simplicity)
\beqa
\bm{J}^{LPGE}_{\perp}=\frac{11\tau e^4B_1 vE^2_0}{30\pi^2\mu b_3(1+\omega^2\tau^2)}\left(\begin{array}{c}
-\cos\theta_i \sin2\alpha\\
2\cos^2\theta_i\cos^2\alpha\\
0
\end{array}\right).\label{LPGCperp}
\eeqa
The dependence with the polarization angle $\alpha$ is the same as in (\ref{LPGC}), but changes the dependence with the incidence angle $\alpha_i$.
\subsection{Circular PGE at wave incidence $\theta_i$}
In this case, the electric field reads
\beqa
\bm{\mathcal{E}}=\frac{E_0}{2}\left(\begin{array}{c}
\cos\theta_i \\
\lambda i\\
\sin\theta_i
\end{array}\right).
\eeqa
The parameter $\lambda=\pm1$ determines if the wave is right or left handed polarized. Inserting this expression now in (\ref{PGEcurrentcomponents}) (again, with $\bm{B}||\bm{b}$), taking care of the complex conjugate part, one gets
\beqa
\bm{J}^{CPGE}=-\frac{\tau e^4B_3 vE^2_0}{60\pi^2\mu b_3(1+\omega^2\tau^2)}\left(\begin{array}{c}
2\lambda \omega \tau\sin\theta_i\\
-3\sin2\theta_i\\
2\lambda\omega\tau\cos\theta_i
\end{array}\right).\label{CPGC}
\eeqa
The most salient consequence of this current is that when the incidence angle is normal to the surface ($\theta_i=0$), a transverse current along the $z$ direction appears. The direction of the current depends on the chirality $\lambda$ of the plane wave.
When $\bm{B}\perp\bm{b}$, we have
\beqa
\bm{J}^{CPGE}_{\perp}=-\frac{11\tau e^4B_1 vE^2_0}{60\pi^2\mu b_3(1+\omega^2\tau^2)}\left(\begin{array}{c}
\lambda \omega \tau\cos\theta_i\\
-\cos^2\theta_i\\
0
\end{array}\right).\label{CPGCperp}
\eeqa
Now the difference with the case $\bm{B}||\bm{b}$ is more pronounced. At normal incidence ($\theta_i=0$), the photoinduced current (\ref{CPGCperp}) is perpendicular to $\bm{b}$, instead of parallel.
\section{Second Harmonic Generation in Dirac semimetals}
\label{sec:allSHG}
Doing the appropriate changes, the SHG current $\bm{J}^{SHG}(2\omega)$ can be read from (\ref{PGEcurrent}):
\beqa
\bm{J}^{SHG}_1(2\omega)&=&-2e^4\frac{\tau}{\alpha_1}\int (d\bm{k})\frac{\partial f_0}{\partial \epsilon_{\bm{k}}}
[(\bm{\Omega\cdot\bm{B}})(\bm{\mathcal{E}}\cdot\bm{v}_0)-\nonumber\\
&-&(\bm{B}\cdot\bm{\mathcal{E}})(\bm{\Omega}\cdot\bm{v}_0)]\bm{\Omega}\times\bm{\mathcal{E}}+c.c.\label{SHGcurrent}
\eeqa 
or, in components, after integrating in $\bm{k}$ ($\bm{B}||\bm{b}$),
\beq
J^{SGH}_a(2\omega)=\left(\frac{e^4B_3b_3\tau v}{30\pi^2 \alpha_1\mu b^2_3}\right)\epsilon_{abc}\Lambda_{bd}\mathcal{E}_c\mathcal{E}_d+c.c.,\label{SHGcurrentcomponents}
\eeq
We se immediately that, because the matrix $\Lambda_{bd}$ differs from the identity matrix, the SHG response in inversion symmetric Dirac semimetals does not vanish when the applied magnetic field has a component parallel to the vector $\bm{b}$. It is important to remember that this expression has been obtained under the assumption of axial symmetry around $\bm{b}$.

When $\bm{B}\perp\bm{b}$,
\beq
\bm{J}^{SHG}(2\omega)=\left(\frac{11e^4\tau v}{30\pi^2 \alpha_1\mu b^2_3}\right)(\bm{B}_{\perp}\cdot\bm{\mathcal{E}})\bm{b}\times\bm{\mathcal{E}}+c.c.\label{SHGxialanomaly}
\eeq

As it happens with the PGE linear in the magnetic field $\bm{B}$ current in the previous section, the SHG current (\ref{SHGcurrentcomponents}) vanishes for the linear Weyl model.
\subsection{Linear SHG at wave incidence $\theta_i$ and polarization angle $\alpha$}

Since the electric field $\bm{\mathcal{E}}$ is real, $\bm{\mathcal{E}}=\bm{\mathcal{E}}^*$, the expressions for the LSHG current are the same as the LPGE:
\begin{subequations}
\beqa
\bm{J}^{LSHG}=\frac{\tau e^4B_3 vE^2_0}{40\pi^2\mu b_3(1+\omega^2\tau^2)}\left(\begin{array}{c}
-\sin\theta_i \sin2\alpha\\
\sin2\theta_i\cos^2\alpha\\
0
\end{array}\right),\label{LSHG}
\eeqa
\beqa
\bm{J}^{LSHG}_{\perp}=\frac{11\tau e^4B_1 vE^2_0}{30\pi^2\mu b_3(1+\omega^2\tau^2)}\left(\begin{array}{c}
-\cos\theta_i \sin2\alpha\\
2\cos^2\theta_i\cos^2\alpha\\
0
\end{array}\right).\label{LSHGperp}
\eeqa
\end{subequations}
The physical meaning is different. They mean that a second harmonic wave will be generated. The precise polarization of this wave can be computed following the reference\cite{BP62}.
\subsection{Circular SHG at wave incidence $\theta_i$}
The electric field we consider is the same as in the previous section for the CPGE: a circularly polarized plane wave of incidence angle $\theta_i$ and :
\beqa
\bm{\mathcal{E}}=\frac{E_0}{2}\left(\begin{array}{c}
\cos\theta_i \\
\lambda i\\
\sin\theta_i
\end{array}\right).
\eeqa
The circularly polarized SHG current is, after using (\ref{SHGcurrentcomponents}),
\beqa
\bm{J}^{CSHG}=\frac{\tau e^4B_3 vE^2_0}{40\pi^2\mu b_3(1+\omega^2\tau^2)}\left(\begin{array}{c}
2\lambda \omega \tau\sin\theta_i\\
-\sin2\theta_i\\
0
\end{array}\right).\label{CSHG}
\eeqa
Again, one needs to solve the Maxwell equations in an interface between vacuum and a Dirac semimetal to compute the non linear Fresnel coefficients\cite{BP62}. We content ourselves with the computation of the expression (\ref{CSHG}).
When ($\bm{B}\perp\bm{b}$), the circular SHG current turns out to be the same as (\ref{CPGCperp}).

\section{Discussion}
\label{sec:discussion}
Here we have computed second non-linear optical responses induced by an external magnetic field $\bm{B}$ in inversion-symmetric and time reversal invariant Dirac semimetals, like Cd$_3$As$_2$ and Na$_3$Bi. The effects responsible of these phenomena are the so-called magnetochiral effect and a version of the chiral magnetic effect. However inversion-symmetric systems in bulk might display second non-linear optical responses due to the lack of inversion symmetry at the surface. Moreover, such surface second responses might depend on the externally applied magnetic field as well\cite{PWS89}. We have to add this unknown optical responses to the ones described in the present work if we want to correctly characterize the SHG wave. However, in the materials mentioned above, the Weyl nodes lie on a high symmetry axis, so tuning adequately the magnetic field, the two effects mentioned above can leave different fingerprints on the photoinduced current, being possible to disentangle them from the effects induced by the presence of the surface.
 
In the case of the photogalvanic effect, the present theory predicts induced dc photocurrents to appear at the surface of a Dirac semimetal when irradiated with linearly polarized light and the magnetic field is collinear with the vector $\bm{b}$ (and this vector pointing perpendicularly to the surface). As we have mentioned earlier, dc photocurrents are allowed in systems that lack inversion symmetry, due to Berry phase effects, however we have to stress that here they appear for linearly polarized light as well, and it is a bulk effect. The benchmarks of these photocurrents are the angular dependence with the magnetic field relative to the vector $\bm{b}$, and that the direction of this photocurrent depends on the angle of the linear polarization.
When circularly polarized light and at normal incidence is used and $\bm{B}$ is perpendicular to $\bm{b}$, the generated dc photocurrent is perpendicular to the surface, and the direction of the current depends on the handedness of the wave.  In this case, we have to keep in mind that all the non-linear currents have been computed for homogeneous electric fields within the bulk. We also have to notice that the system, although displaying topological features, is still a metal, so the electromagnetic field is attenuated when passing through the system. It means that the assumption of homogeneous electric fields at finite $\omega$ is not even approximate, and the computation of the induced current has to be done self consistently with the electric field profile. This implies that the lack in inversion in the profile of the electric field might in turn induce inversion symmetry breaking effects  in inversion-symmetric systems, in the same way the surface modifies the optical activity in the linear regime\cite{AY73}. These effects are beyond the scope of the present analysis. Alternatively, one can think in thin film samples of Dirac semimetals, but in this case, it is known that the quantum confinement induced in these geometries can again induce inversion symmetry breaking effects\cite{MO10}. It would be interesting to perform a proper computation taking into account this spatial dependence. We leave it for future research.

In this work, we have considered the effect of the magnetic field through the semiclassical equations of motion, taking into account the effect of the orbital magnetic moment of the wavepacket on the dispersion relation $\varepsilon(\bm{k})$. In addition, the magnetic field can also modify the dispersion relation through a Zeeman coupling $\bm{J}\cdot \bm{B}$ that takes into account the specific angular momentum of the atomic orbitals that compose the relevant Bloch state. It has been discussed recently\cite{CBW16} that the inclusion of this Zeeman term might lift the degeneracy of the Weyl nodes for each Kramers partner in a non trivial way, depending on the direction of the magnetic field. In our case, since we study optical effects linear in the magnetic field, we are assuming relatively small fields, so we expect the effects of the Zeeman term to be weak enough, and the results presented here to be valid.

Finally, we have focused only in non linear optical responses depending linearly on the magnetic field. It is expected that quadratic (and higher) dependence in $\bm{B}$ to appear in the current associated to $f_2$. In this case, these quadratic terms are expected to appear directly when the linear model for Dirac semimetals is considered\cite{MZO16}, since they are quadratic in $\bm{\Omega}$ or $\bm{m}$, being the terms depending on $\bm{b}$ sub-leading. Also, the use of semiclassical kinetic methods might reduce the range of frequencies where this theory is applicable to frequencies $\omega<2\mu$, thus limiting in principle the experimental conditions to observe these effects. In the reference \cite{MZO16} it has been pointed out that similar results to the ones from the semiclassical kinetic theory are obtained when a quantum treatment including interband processes is used. This means that the results obtained in the present work will be obtained using these methods as well, and the range of qualitative applicability should be larger. Cd$_{3}$As$_{2}$ has been experimentally probed in ARPES experiments up to Fermi energies around 200 meV\cite{LGA15}. This means that the present theory is accurate for irradiated light with frequencies corresponding to (mid/near) infrared frequencies.
\section{Acknowledgements} 
\label{sec:acknoledgements}
The author acknowledges financial support from the European Union structural funds and the Comunidad de Madrid MAD2D-CM Program (S2013/MIT-3007) and MINECO (Spain) Grant No. FIS2015-73454-JIN.
\appendix
\section{Computation of $J^{(2,1)}_a$}
In order to compute the momentum integral in (\ref{PGEcurrentcomponents}) we need expressions for the velocity $v^0_a$ and the Berry curvature $\Omega_{a}$. The Hamiltonian (\ref{Ham0}) can be written as $H(\bm{k})=\bm{\sigma}\cdot\bm{d}(\bm{k})$. We will exploit the fact that components of the vector $\bm{d}$ only depend on the same component of $\bm{k}$: $d_{a}(\bm{k})=d_a(k_a)$. Employing this generic form for the Hamiltonian, we can write, for the conduction band described by the eigenstate $\ket{+}$ ($\epsilon_{\bm{k}}=|\bm{d}|$):
\beq
v^0_a=\frac{\partial \epsilon_{\bm{k}}}{\partial k_a}=\left(\frac{\partial d_a}{\partial k_a}\right)\frac{d_a}{\epsilon_{\bm{k}}},
\eeq
and
\beq
\Omega_a=i\varepsilon_{abc}\langle \partial_b + | \partial_c + \rangle=-\frac{1}{2\varepsilon^3_{\bm{k}}}\left(\frac{\partial d_a}{\partial k_a}\right)^2 d_a.
\eeq
We then use these expressions in 
\beqa
J^{(2,1)}_a&=&-2e^4\frac{\tau}{\alpha_1}\epsilon_{abc}\int (d\bm{k})\frac{\partial f_0}{\partial \epsilon_{\bm{k}}}\Omega_b\Omega_r\cdot\nonumber\\
&&\cdot[B_r v^{0}_d-B_dv^{0}_r]\mathcal{E}^*_c\mathcal{E}_d+c.c,
\eeqa
assuming zero temperature, $\frac{\partial f_0}{\partial \epsilon_{\bm{k}}}=-\delta(\mu-\epsilon_{\bm{k}})$. Since the Hamiltonian $H(\bm{k})$ is not isotropic, we use cylindrical coordinates:$k_1=k\cos\theta$, $k_2=k\sin\theta$, and $k_3$, so $(d\bm{k})=\frac{1}{8\pi^3}d k_3dk kd\theta$. Since $\epsilon_{\bm{k}}=\sqrt{v^2k^2+d^2_3(k_3)}$, we can write the integral with the Dirac delta as
\beqa
 &&\frac{1}{8\pi^3}\int d k_3dk kd\theta\delta(\mu-\epsilon_{\bm{k}})(...)= \nonumber\\
 &+&\frac{1}{8\pi^3}\int d k_3d\theta dk k\frac{\mu}{v^2 k_+}\delta(k-k_+)\Theta(k_+)(...),
\eeqa
with $k_+=\sqrt{\frac{\mu^2}{v^2}-d^2_3(k_3)}$. 
Then the current reads, for the magnetic configuration $\bm{B}=(0,0,B_3)$:
\beqa
J^{(2,1)}_a&=&-e^4\frac{\tau\mu B_3}{4\pi^3\alpha_1v^2}\epsilon_{abc}\int d k_3d\theta d k\delta(k-k_+)\Theta(k_+)\nonumber\\
&&\Omega_b\cdot[\Omega_3v^{0}_d-\delta_{d3} \Omega_rv^{0}_r]\mathcal{E}^*_c\mathcal{E}_d+c.c.\label{component11}
\eeqa
The set of components that give non zero contributions after integrating over $\theta$ are $(b=1,d=1)$, $(b=2,d=2)$, and $(b=3,d=3)$. By symmetry, the integral for $(b=1,d=1)$ will be the same as for $(b=2,d=2)$:
\beqa
J^{(2,1)}_a&=&-e^4\frac{\tau\mu B_3}{4\pi^3\alpha_1v^2}\epsilon_{a1c}\int d k_3d\theta d k\delta(k-k_+)\Theta(k_+)\nonumber\\
&&\Omega_1\Omega_3v^{0}_1 \mathcal{E}^*_c\mathcal{E}_1+c.c,\label{component22}
\eeqa
while for $(b=3,d=3)$ we have
\beqa
J^{(2,1)}_a&=&e^4\frac{\tau\mu B_3}{4\pi^3\alpha_1v^2}\epsilon_{a3c}\int d k_3d\theta d k\delta(k-k_+)\Theta(k_+)\nonumber\\
&&\Omega_3\cdot[\Omega_1v^{0}_1+\Omega_2v^{0}_2]\mathcal{E}^*_c\mathcal{E}_3+c.c.\label{component33}
\eeqa
Now the integrals can be easily done by using the definitions of $\Omega_a$ and $v^0_a$, and substituting $k$ by $k_+$ and $\epsilon_{\bm{k}}$ by $\mu$. The step function $\Theta(k_+)$ gives the integration limits for $k_3$: $\Theta(k_+)$ implies that the integral is limited by the condition $\frac{\mu^2}{v^2}-d^2_3(k_3)=0$, so we write $k_3$ as a function of $\mu$. Using the definition of $m$ and $\beta$ in terms of $b_3$ and $v$ in the Hamiltonian (\ref{Ham0}), the integration limits come from the solution of the algebraic equation
\beq
\left(1-\frac{k^2_3}{b^2_3}\right)^2=\frac{4\mu^2}{v^2 b^2_3}\equiv\delta^2,
\eeq
with solutions
\beq
k_3=\pm b_3\sqrt{1\pm \delta}.
\eeq
We then have two regions: $0<\delta<1$, where we have four solutions to this equation. This region corresponds to the situation when the Fermi level $\mu$ lies below the Van Hove point, and we have two independent cones. The integral in $k_3$ is split into two parts, with integration limits $-\sqrt{1+\delta}$, $-\sqrt{1-\delta}$, and $\sqrt{1+\delta}$, $\sqrt{1+\delta}$.
The second region corresponds to $\delta>1$. Here we have only two solutions, and this region corresponds to Fermi levels above the Van Hove point where there is a single Fermi surface. The integration limits are $-\sqrt{1+\delta}$ and $\sqrt{1+\delta}$. We will focus on the first region, that allows to compare with the result obtained with the model of linear Weyl fermions ($\mu/ b_3\ll 1$).

The integral in (\ref{component11}) reads, after integrating in $k$:
\beqa
&&\int d k_3d\theta\Omega_1\Omega_3v^{0}_1=\nonumber\\
&=&\int dx d\theta \frac{4 v (1-x^2)(\delta^2-(1-x^2)^2)\cos^2\theta}{\delta^7 b^2_3},
\eeqa
where we have defined $x=k_3/b_3$ and used that $\delta=2\mu/vb_3$. The integral limits for $x$ are the ones described above for the first region of parameters. Now the integrals in $\theta$ and $x$ are trivial. We can expand the result in powers of $\delta$, perform the same steps for (\ref{component33}) obtaining the desired result, equation (\ref{PGEcurrentcomponents}). 
Also this integral can be performed in the opposite limit, $\delta\gg1$, well above the Van Hove point, where the Fermi surface is simply connected and the notion of chiral states is not well defined, and obtain
\beqa
J^{PGE}_a&\simeq&\left(\frac{8e^4B_3b_3\tau v}{45\pi^2 \alpha_1\mu b^3_3}\right)\left(\frac{v b_3}{2\mu}\right)^{\frac{3}{2}}\epsilon_{abc}\Lambda_{bd}\mathcal{E}^*_c\mathcal{E}_d+\nonumber\\
&+&c.c.\label{PGEcurrentcomponents2}
\eeqa
We can repeat the same steps for a magnetic field perpendicular to $\bm{b}$ (for instance, having only $B_1$). The expression (\ref{component11}) reads in this case 
\beqa
J^{(2,1)}_a&=&-e^4\frac{\tau\mu B_1}{4\pi^3\alpha_1v^2}\epsilon_{abc}\int d k_3d\theta d k\delta(k-k_+)\Theta(k_+)\nonumber\\
&&\Omega_b\cdot[\Omega_1v^{0}_d-\delta_{d1} \Omega_rv^{0}_r]\mathcal{E}^{*}_c\mathcal{E}_d+c.c.\label{component11perp}
\eeqa
Now, only the values of $(b,d)$ that gives a non-vanishing contribution are only $(b=3,d=1)$, obtaining, in the limit $\mu\ll b_3$:
\beq
J^{(2,1)}_a=-\left(\frac{11e^4 v\tau}{30\pi^2\alpha_1\mu b_3}\right)\epsilon_{a3c}B_1\mathcal{E}_{1}\mathcal{E}^{*}_c+c.c.
\eeq
If one alternatively choose the magnetic field to have only a $B_2$ component, the result is:
\beq
J^{(2,1)}_a=\left(\frac{11e^4 v\tau}{30\pi^2\alpha_1\mu b_3}\right)\epsilon_{a3c}B_2\mathcal{E}_{2}\mathcal{E}^{*}_c+c.c.
\eeq
From these two expressions, and remembering that $\bm{b}=b_3\hat{\bm{z}}$ in the particular model used, one can conclude that the generic form of the current $\bm{J}^{(2,1)}$ is
\beq
\bm{J}^{(2,1)}\propto (\bm{B}_{\perp}\cdot\bm{\mathcal{E}})\bm{b}\times\bm{\mathcal{E}}^*,
\eeq
when $\bm{B}\perp\bm{b}$.
\section{Linear conductivity}
Although not of primary importance in the present work, we compute in this appendix the linear conductivity up to first order in the magnetic field. This will eventually needed when computing the effect of the non-linear terms on the transmitted wave, since the Maxwell equations for the transmitted and reflected second harmonic wave depends on the electromagnetic field for the transmitted fundamental harmonic\cite{BP62}.

The terms of the current linear in the electric field, and up to linear order in the magnetic field $\bm{B}$ that survive to the sum over Kramers partners are:
\beqa
&&\bm{J}^{(1)}_1(\omega)=-\frac{\tau}{\alpha_1} e^2\int (d\bm{k})\frac{\partial f_0}{\partial \epsilon_{\bm{k}}}\bm{v}_0(\bm{v}_0\cdot \bm{\mathcal{E}})+\nonumber\\
&+&\frac{\tau^2}{\alpha^2_1}e^3\int (d\bm{k})\frac{\partial f_0}{\partial \epsilon_{\bm{k}}}\bm{v}_0(\bm{v}_0\times \bm{B})\cdot\frac{\partial (\bm{v}_0\cdot\bm{\mathcal{E}})}{\partial \bm{k}}.
\label{1ordercurrent}
\eeqa
The first term corresponds to the standard linear conductivity in absence of the magnetic field, while the second, linear in $\bm{B}$, gives the classical Hall current:
\beqa
J^{(1)}_a(\omega)&=&\left(\frac{e^2\tau \mu^2}{3\pi^2\alpha_1v}\right)\delta_{ab}\mathcal{E}_b+\nonumber\\
&+&\left(\frac{e^3\tau^2 v \mu}{6\pi^2\alpha^2_1}\right)\epsilon_{abc}B_c\mathcal{E}_b.
\eeqa
This expression is obtained in the limit $\mu\ll vb_3$, and is the same result that one obtains if the model of linear Weyl fermions is used. These expressions have sub-leading terms that depend on $b_3$. Also, one can easily compute these expressions in the opposite limit $\mu\gg vb_3$ as it is done in ref.\cite{A16}. We will not do that here.

\end{document}